\documentclass[a4paper]{article}

\usepackage[T1]{fontenc} 
\usepackage{float}       
\usepackage{helvet}      
\usepackage{rsc}         
\usepackage{setspace}    
\usepackage{epsfig}
\usepackage{graphicx}
\usepackage{leqno}

\AtBeginDocument{\doublespacing}

\newfloat{chart}{htbp}{loc}
\newfloat{graph}{htbp}{loh}
\newfloat{scheme}{htbp}{los}

\usepackage[version=3]{mhchem} 

\begin{document}

\title{Capillary micromechanics: Measuring the elasticity of microscopic soft objects}
\author{}
\date{}



\maketitle

Hans M. Wyss$^{1,2,\ast}$, Thomas Franke$^{1,3}$, Elisa Mele$^{1,4}$, David A. Weitz$^{1}$ \\

$^1$ Harvard University, Department of Physics \& SEAS, Cambridge (MA), USA\\
$^2$ Eindhoven University of Technology, WTB/MaTe \& ICMS, Eindhoven, the Netherlands\\
$^3$ University of Augsburg, Experimental Physics I, Microfluidics  Group, Universitaetstr. 1, D-86159 Augsburg, Germany\\
$^4$ National Nanotechnology Laboratory of INFM-CNR, University of Lecce, Via Arnesano, I-73100, Lecce, Italy\\

$^\ast$  Correspondence should be addressed to H.M.Wyss@tue.nl

\begin{abstract}
\noindent 
We present a simple method for accessing the elastic properties of microscopic deformable particles. This method is based on measuring the pressure-induced deformation of soft particles as they are forced through a tapered glass microcapillary. It allows us to determine both the compressive and the shear modulus of a deformable object in one single experiment. Measurements on a model system of poly-acrylamide microgel particles exhibit excellent agreement with measurements on bulk gels of identical composition. Our approach is applicable over a wide range of mechanical properties and should thus be a valuable tool for the characterization of a variety of soft and biological materials.
\end{abstract}

\newpage

\subsection*{Introduction}
Soft mesoscopic particles are remarkably common in biological systems, industrial processes and everyday products such as foods, household products, pharmaceutics or cosmetics\cite{malmsten2006soft,stokes2008rheology,jeong2002lessons,kasza2007cell}. 
However, the mechanical behavior of systems that consist of deformable objects is still surprisingly poorly understood. At low particle concentrations soft objects exhibit a viscoelastic response similar to hard spheres; however, as the concentration increases they can behave in a drastically different way. Because they are deformable, highly packed suspensions of soft particles exhibit a comparably much lower viscosity as the particles can deform and shrink in response to an increase in concentration\cite{adams2004influence,Borrega:1999p6058,Ketz:1988p2084,Mattsson:2009p6046}. Soft microgel particles thus do not exhibit the dramatic, critical-like increase in viscosity that is seen in hard spheres\cite{Ketz:1988p2084,Saunders:1999p2086} as their volume fraction approaches close packing; instead, their viscosity increases more gradually, depending on the softness of the individual particles\cite{Borrega:1999p6058,Mattsson:2009p6046}.

This illustrates the importance of mechanical characterization methods for microscopic soft objects. Such methods could also be useful for measuring the mechanical behavior of additives used in the food and drug industry, as well as for the characterization of cells and other biological materials.
While existing techniques such as atomic force microscopy (AFM)\cite{AFM_microgels_Lyon_2007,Matzelle:2003p6075,AFM_Rotsch_2000,AFM_Hassan_1998,tagit2008probing} or micropipette aspiration\cite{micropipette_aspiration_Kwok_Evans} can be used to characterize a mechanical response at small scales, these methods are often difficult to carry out as they require localizing each particle under a microscope. In addition, they probe a highly localized response at the surface of a soft particle but do not readily provide information on the elastic response of the entire particle.

In this article we present a simple and direct method for characterizing the mechanics of microscopic soft objects. Our approach directly characterizes the elastic properties of entire soft objects. We use a microfluidic setup where particles are deformed in tapered microcapillaries; direct imaging of their deformation with an optical microscope enables us to characterize both the compressive and the shear modulus of a soft particle in a single experiment.

\begin{figure}[!h] 
\centering
\includegraphics[width=\columnwidth]{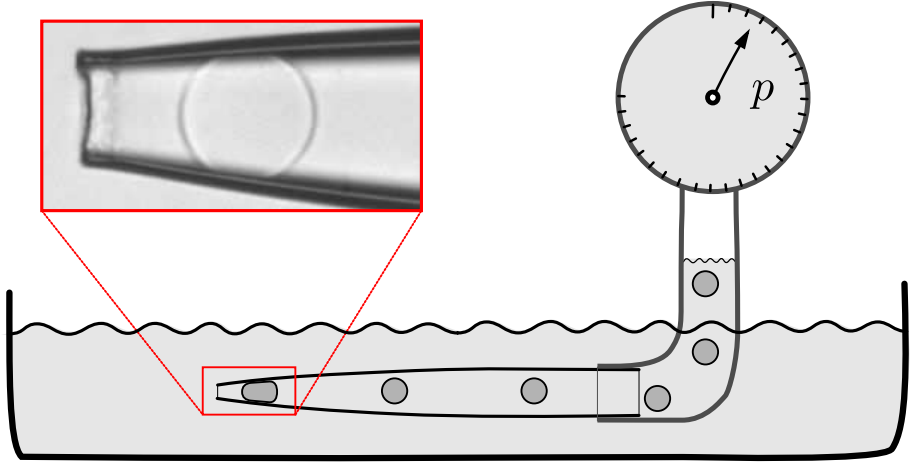}
\caption{\emph{Experimental setup:} \ 
A dilute suspensions of soft particles flows towards the tapered tip of a microcapillary as a result of an applied pressure difference. For low pressure values $<10^3$ Pa we exploit gravity and use the height difference of the sample reservoir and outlet of the microcapillary tip to apply a hydrostatic pressure. Higher pressures can be achieved by connecting the sample reservoir to a pressure regulator. As a single particle blocks the flow of fluid, the particle deforms until the pressure-induced external stresses are balanced by its internal elastic stress. The pressure-dependence of the shape and size of the particle is thus a direct measure of its elastic properties.}
\label{exp_setup}
\end{figure}

\subsection*{Materials and Methods / Experimental}
We flow a dilute suspension of particles through a glass capillary by applying a pressure difference $p$ between the inlet and the outlet of the capillary, as shown schematically in Fig.\ \ref{exp_setup}. The capillary is tapered towards the tip, where the inner diameter at tip is smaller than the size of a single particle. Therefore, a single particle eventually clogs the capillary and blocks further flow of the fluid. In this situation the entire pressure difference falls off across the length of the particle; as a result, an external stress is applied to the particle, causing it to deform. The particle is also subjected to a stress in the radial direction, due to the normal forces exerted on the particle by the capillary walls; the ratio between this radial stress and the applied pressure is directly determined by the taper angle of the glass capillary.  In equilibrium, when the particle no longer moves or changes shape, these externally applied stresses must match the internal stresses resulting from the elastic deformation of the particle.
This balance between the externally applied stresses and the internal elastic stresses enables us to directly derive the elastic properties of a particle by following its deformation as $p$ is increased.
The setup can be used to apply a wide range of pressures on the particle, ranging from 1 Pa  up to $10^6$ Pa; in addition, the radial stresses can be tuned by using capillaries with different taper angles. Hence, the technique can be employed to make measurements over a wide range of elastic material properties.

\begin{figure}[!h] 
\centering
\includegraphics[width=\columnwidth]{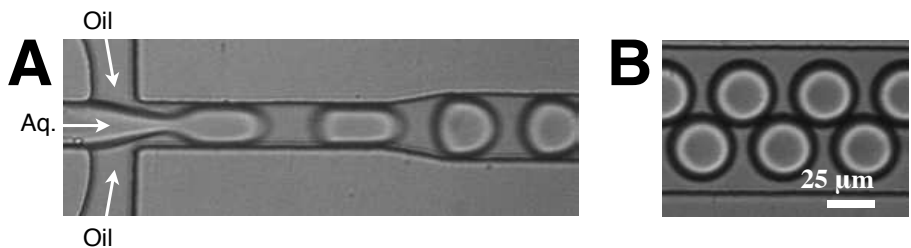}
\caption{\emph{Making poly-acrylamide particles in microfluidics} 
(A) Flow focussing device to produce a water-in-oil emulsion. The aqueous phase, injected in the middle channel, contains acrylamide monomer, the cross-linker BIS-acrylamide, and the catalyst ammonium persulfate. The oil phase is injected in the two outer channels and contains a surfactant stabilizing the drops as well as Tetramethylethylendiamin (TEMED), which acts as an initiator for the polymerization of acrylamide. The device is fabricated from a PDMS elastomer using standard soft lithography techniques\cite{Xia_Whiteside_Angewandte_1998}; the channel walls are subsequently treated to be hydrophobic, thereby enabling a stable production of water drops within a continuous phase of oil\cite{shah2008fabrication}. (B) A picture taken further downstream illustrates that the produced drops exhibit a highly uniform size distribution and they do not coalesce.}
\label{microfluidics}
\end{figure}

To verify our micromechanical method, we use the most general type of elastic material, which is both deformable and compressible. We use microgel particles which display both a shear resistance and a compressibility\cite{Saunders:1999p2086,pelton2000temperature}.
The particles consist of a sparse, crosslinked polymer network of poly-acrylamide, with a background fluid of water; the elastic properties of this system can be conveniently tuned by adjusting the concentration of polymer and of the cross-linker during synthesis.
To achieve a uniform size distribution of particles we produce the particles using microfluidics\cite{squires2005microfluidics}\cite{shah2008fabrication}. We make uniformly sized aqueous drops using a microfluidic flow-focussing device, as shown in Fig.\ \ref{microfluidics}(A). The drops are formed from the aqueous inner phase, which contains the acrylamide monomer and the cross-linker BIS-acrylamide. The outer phase consists of HFE-7500 fluorocarbon oil and contains 5\% vol/vol 1H,1H,2H,2H-perfluoro-1-octanol (Sigma),
stabilized by 1.8 wt \% of the fluorosurfactant ammonium carboxylate of DuPont (Krytox 157) a surfactant that stabilizes the emulsion drops, as well as the polymerization initiator Tetramethylethylendiamin (TEMED). The drops are collected in a vial, where they remain overnight as the polymerization takes place. Subsequently, we wash the particles with a series of centrifugation and dilution steps to remove the surfactant from the surface and to disperse the particles in a background fluid of water. Because the particles are no longer encapsulated in drops, they can now swell to a size $V$, which exceeds the initial drop size $V_0$; this equilibrium size depends on the initial monomer concentration $c_p$ and cross-link density present in the drops before the washing begins. We prepare a series of samples, keeping the ratio of cross-linker to monomer constant at 2.6\% by weight, while varying the monomer concentration in a range from 4\% to 10\% weight fraction. 

We use these particles in our capillary device to determine their elastic properties and validate our micromechanical measurement technique. To minimize friction between the particles and the wall, we treat the glass surface by flowing a solution of bovine serum albumin (BSA) through the channel, an approach frequently used to prevent cells and proteins from sticking to glass surfaces in biological studies\cite{delamarche1998microfluidic,hou2009deformability,arima2007effects} ; it also works well for our polyacrylamide microgel particles. Subsequently, we flush the device with water and then flow a dilute suspension of particles through the capillary by applying a pressure difference $p$ between the inlet and the outlet, as schematically illustrated in Fig.\ \ref{exp_setup}. We ensure that at the moment a particle first clogs the capillary, the applied pressure is at the lowest value in the range of pressures we want to access in the experiment. 

\subsection*{Results}
In a typical experiment we start from this low pressure and subsequently increase $p$ in small steps, while simultaneously capturing the change in shape and volume of the particle using a optical microscope equipped with a digital camera. 
At each pressure we wait until the particle shape no longer changes; for the systems studied a time step of of 60\ s is sufficient to achieve this equilibration, ensuring that we characterize the elastic properties in the zero frequency limit.

\begin{figure}[h] 
\begin{center}
\includegraphics[width=0.8 \columnwidth]{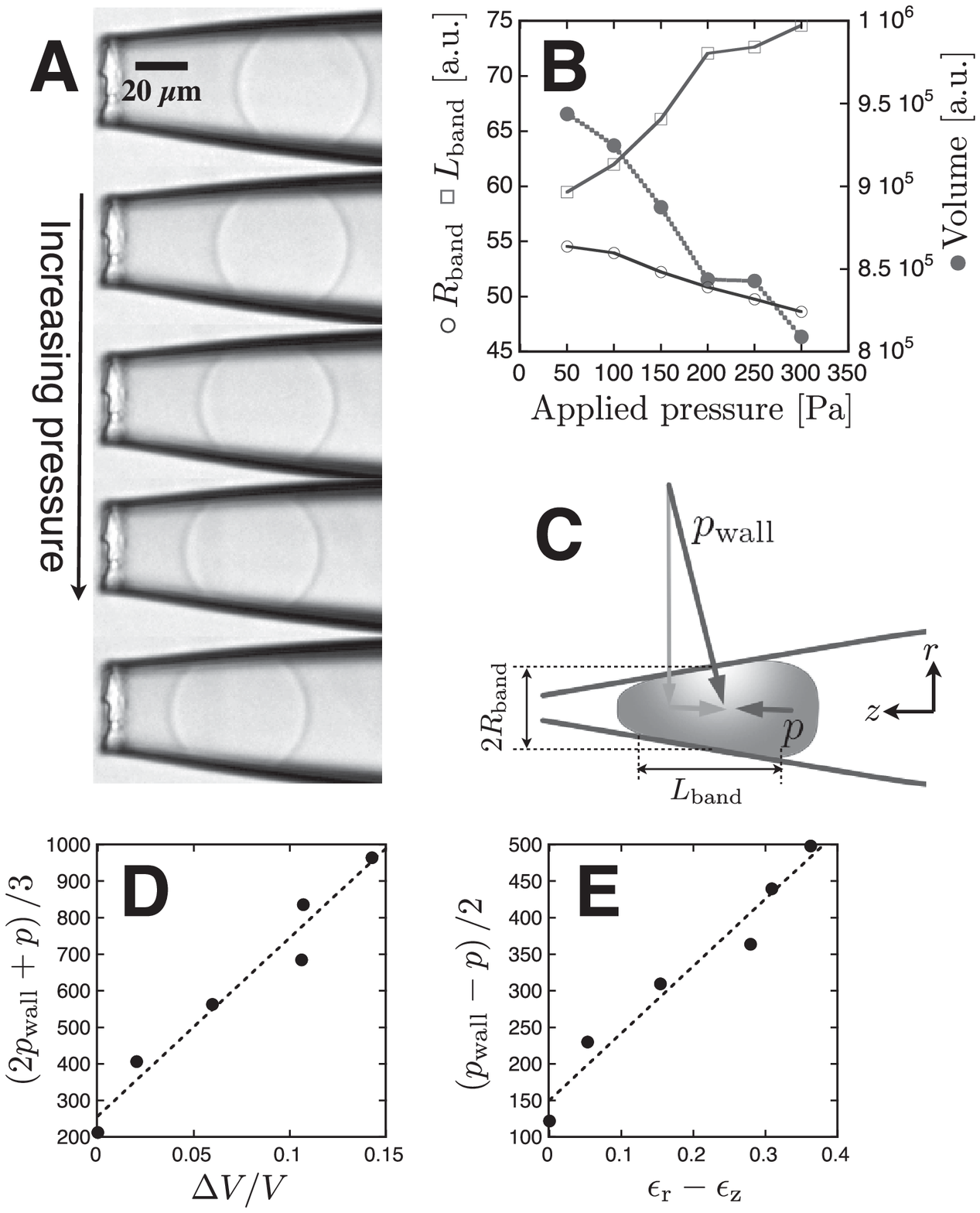}
\end{center}
\caption{ \label{typical_exp} \emph{A typical capillary micromechanics experiment} \ 
(A) Series of images of a microgel particle being deformed as the applied pressure increases; from top to bottom: $p=$100\ Pa, 150\ Pa, 200\ Pa, 250\ Pa, 300\ Pa.
(B) Change of the particle geometry. The contact area where the particle surface is in contact with the glass wall has the shape of a tapered band, with average radius $R_\mathrm{band}$ and width $L_\mathrm{band}$. As the applied pressure $p$ increases, $R_\mathrm{band}$ (open circles) decreases and $L_\mathrm{band}$ (open squares) increases. The volume of the particle decreases, indicating that the particle has a finite compressive modulus. \ 
(C) Schematic of deformation. In the absence of friction the $z$-component of all contact forces between the wall and the particle balance the direct force due to the pressure difference $p$ between the front and the rear of the particle. $L_\mathrm{band}$ and $R_\mathrm{band}$ are the length and the radius of the contact band between the particle and the wall of the glass device.\ 
(D) Compressive stress as a function of the volumetric strain; the dashed line is a linear fit to the data;  the compressive modulus of the particle is given by the slope of this curve, $K \approx 4.5\ $kPa. \  
(E) Differential stress $(p_\mathrm{wall}-p)/2$ as a function of $\epsilon_\mathrm{r}-\epsilon_\mathrm{z}$; the elastic shear modulus is the slope of the linear fit curve, $G \approx 0.8\ $kPa.
}
\end{figure}

Typical results are shown in the series of images in Fig.\ \ref{typical_exp}(A), where a particle with $c_p$=6\% is shown under an applied pressure difference that increases from $p=$100\ Pa in the top image to 300\ Pa in the bottom image. 
The particle clearly changes its shape; as the pressure is increased it becomes more elongated and is compressed along the radial direction as it moves closer to the tip of the capillary. To quantify this shape change we analyze our microscope images and extract geometrical information such as the volume $V$ of the particle, its length, as well as its width. 

The surface of the particle that is in contact with the glass walls has the shape of a tapered band with circular cross section; we use the length $L_\mathrm{band}$ and the average radius $R_\mathrm{band}$ of this band as a measure of the length and the radius of the particle, respectively, shown in Fig.\ref{typical_exp}(C). As the pressure is increased,  $L_\mathrm{band}$ increases, while $R_\mathrm{band}$ decreases continuously, as shown in Fig.\ \ref{typical_exp}(B). Moreover, a significant change in volume is also observed; the particle is compressed as the applied pressure is increased, as shown in Fig.\ \ref{typical_exp}(B).

The dependence of these parameters on the applied pressure is a direct consequence of the elastic properties of the particles and can thus be employed to compare the properties of different particles relative to each other. While the relative comparison of elastic properties is straightforward, a more detailed analysis is necessary to adequately quantify the elastic properties of the particles and to make results comparable to generally used elastic moduli. Moreover, a detailed analysis allows us to account for the full elastic response of the particles.

In any isotropic elastic material two independent moduli such as the compressive and the shear modulus, $K$  and $G$ , are sufficient to describe the full elastic behavior\cite{Hunter_mechanics_book}. Here we use the combined information on the changes in shape and volume as well as the stresses applied to the particle to obtain both moduli from a single experiment. 

The stress on the particle due to external forces is given by the pressure difference applied to the particle. In a equilibrium situation, this stress must be balanced by the elastic stress, which is a function of the particle's deformation, as well as the moduli $K$ and $G$. 
The externally applied stresses are directly proportional to the applied pressure difference. As our particles consist of a polymer network with a incompressible background fluid, the absolute pressure has no effect on the stress exerted on the particle itself; only the pressure difference applied to the particle is relevant and we can neglect the effect of the atmospheric background pressure. Due to the porous nature of the particles, fluid will continue to flow through the pores of the polymer network, thereby exerting viscous drag forces on the polymer chains. We therefore equate the stress along the longitudinal direction of the capillary with the applied pressure difference; $\sigma_z = p$. 
The pressure exerted by the wall of the capillary on the particle can be derived by assuming the absence of static friction between the particle and the wall. 
Microgel particles are known to exhibit very low static friction at interfaces, which is also the reason why wall slip is often observed in rheological measurements on these materials.
The influence of friction can be tested experimentally by examining the degree of syneresis between increasing and decreasing pressure steps. For the system studied here we find only a small degree of syneresis, indicating that static friction is not important. However, for other soft objects static friction may play a significant role and should be taken into account in analyzing the data.

In the absence of static friction, the longitudinal component of all forces from the wall must balance the force exerted in that direction through the applied pressure difference, as shown schematically in Fig.\ \ref{typical_exp}(C). The longitudinal component of the force acting on an area element $\Delta{A}$ on the surface is thus given by $p_\mathrm{wall}  \Delta A \sin(\alpha)$. Integrating these forces over the total contact area between the particle and the wall yields a total longitudinal force $F_{\parallel \mathrm{wall} } \approx 2 \pi R_\mathrm{band} L_\mathrm{band} \sin(\alpha)$. In equilibrium, this force must balance the longitudinal force $F_\parallel=p \pi R_\mathrm{wall}^2$ due to the applied pressure difference $p$. As a result, the average wall pressure is derived as

\begin{equation}
p_{\mathrm{wall}} = \frac{2}{\sin(\alpha)}\frac{R_{band}}{L_{band}} p , \label{eq:pwall}
\end{equation}

where $\alpha$ is the taper angle of the capillary, while $R_\mathrm{band}$ and $L_\mathrm{band}$ are the radius and length of the band around the particle that is in contact with the glass wall. Thus, assuming a uniform stress distribution within the particle, the stress in the radial  direction is directly given by the wall pressure, $\sigma_r=p_\mathrm{wall}$ and in the longitudinal direction it is given by the applied pressure difference, $\sigma_z=p$. We use a cylindrical coordinate system, where the $z$-direction is the central axis of the capillary, as shown in Fig.\ref{typical_exp}(C).

In equilibrium, when the particle is no longer moving, the stress exerted by external forces must be balanced by the internal elastic stresses within the particle, which, for an isotropic material can be written as a function of $K$, $G$, as well as the three-dimensional strain deformations, as\cite{Hunter_mechanics_book, Macosko_Rheology_book}

\begin{eqnarray}
\sigma_\mathrm{r} = K \left( 2 \epsilon_\mathrm{r} + \epsilon_\mathrm{z} \right) + \frac{2}{3} G \left( \epsilon_\mathrm{r} - \epsilon_\mathrm{z} \right) \\
\sigma_\mathrm{z} = K \left( 2 \epsilon_\mathrm{r} + \epsilon_\mathrm{z} \right) - \frac{4}{3} G \left( \epsilon_\mathrm{r} - \epsilon_\mathrm{z} \right) 
\end{eqnarray}

Balancing this internal elastic stress with the externally applied stress implies $\sigma_\mathrm{r}=p_{\mathrm{wall}}$ and $\sigma_\mathrm{z}=p$. By solving these equations for $K$ and $G$ the elastic moduli are directly expressed as a function of the applied pressures and strain deformations $\epsilon_\mathrm{r}$ and $\epsilon_\mathrm{z}$:

\begin{equation} {K = \frac{  \frac{1}{3} (2 p_\mathrm{wall} + p )}{2 \epsilon_\mathrm{r} + \epsilon_\mathrm{z}} }  \label{eq:k}
\end{equation}

\begin{equation}
G  = \frac{ \frac{1}{2} (p_\mathrm{wall} - p)}{\epsilon_\mathrm{r} - \epsilon_\mathrm{z} }  . \label{eq:g}
\end{equation}

Both expressions can be rationalized in terms of a elastic stress in response to a strain deformation characteristic of the mode of deformation probed. 
The compressive modulus $K$ quantifies resistance of the material to volume change; the relevant stress is the hydrostatic pressure, which is the average of the stresses along the principal axes, thus $\sigma_\mathrm{compr} = (2 p_\mathrm{wall} + p) / 3$, whereas the characteristic strain is the volumetric strain $\Delta V / V \approx 2 \epsilon_r + \epsilon_z$. The compressive modulus is thus determined from the slope of $\sigma_\mathrm{compr.}$ as a function of $\Delta V / V$, as shown for a typical experiment in Fig.\ \ref{typical_exp}(D).  

The shear modulus $G$ quantifies the resistance of the material to a shape deformation; the relevant stress is the difference of the stress in the longitudinal and the radial direction, $\sigma_\mathrm{shear} = (p_\mathrm{wall} - p) / 2$, whereas the characteristic strain characterizes the shape change; it is the difference of the strains in the longitudinal and the radial direction $\epsilon_\mathrm{shear}=\epsilon_r - \epsilon_z$. The shear modulus is thus determined from the slope of $\sigma_\mathrm{shear}$ as a function of $\epsilon_\mathrm{shear}$, as shown for a typical experiment in Fig.\ \ref{typical_exp}(E).

\begin{figure}[h!] 
\begin{center}
\includegraphics[width=0.8 \columnwidth]{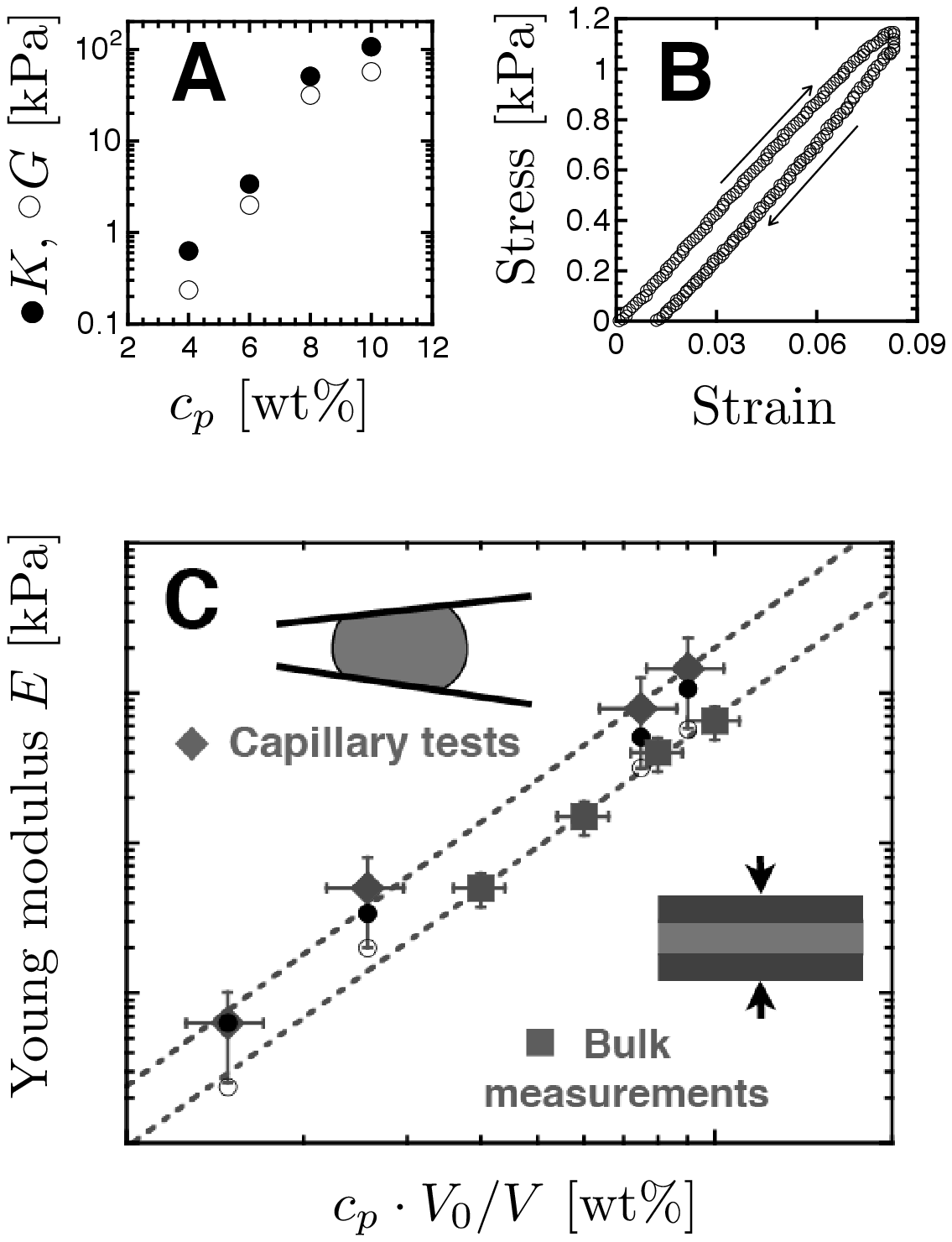}
\end{center}
\caption{ \label{Mechanics_vs_cp} \emph{Mechanical properties as a function of polymer concentration:} \ 
(A) Compressive modulus $K$ (solid circles) and shear modulus $G$ (open circles) as a function of the polymer concentration $c_\mathrm{p}$. \ 
(B) Typical data from a bulk compressive mechanical test. The curve shown is for a sample with 6\% polymer concentration; the arrows indicate the order of data as the sample is compressed and then immediately decompressed again at a rate of 0.1 mm/min.
(C) Young's modulus $E$ from compressive mechanical tests on bulk poly-acrylamide gels (green squares) and for microgel particles (red diamonds) as derived from the compressive modulus $K$ (solid circles) and the shear modulus $G$ (open circles). To account for the swelling of the particles in water, the polymer concentration is corrected by the volume ratio $V_0/V$, where $V_0$ is the volume of the particles as they are formed in the microfluidic devices, and $V$ is the equilibrium particle volume after swelling in water.}
\end{figure}

We thus obtain both the compressive and the shear modulus of a particle from a single experiment. The results for the different poly-acrylamide samples are shown in Fig.\ \ref{Mechanics_vs_cp}(A) as a function of the weight fraction $c_\mathrm{p}$ of monomer in the microfluidic synthesis of the particles. The magnitude of both $K$ and $G$ increases continuously as $c_\mathrm{p}$ is increased, with the compressive modulus always remaining larger than the shear modulus over the entire range of polymer concentrations studied; the ratio between $K$ and $G$ remains approximately constant at $K/G\approx 2-3$ for all $c_\mathrm{p}$, a behavior that is typical for many polymer gels. Our data corresponds to Poisson ratios $\nu=0.28 \pm 0.04$, a range of values that is in agreement with previous measurements conducted on macroscopic polyacrylamide gels \cite{Li:1993p6077} that match the chemical composition of our gel particles.

\emph{Error analysis}. The limited resolution of the microscope and the digital image analysis as well as the accuracy of the applied pressure introduce errors in the determination of $K$ and $G$ via equations (\ref{eq:k}), (\ref{eq:g}) and (\ref{eq:pwall}). To analyze the propagation of these errors, we elucidate the dependence of $K$ and $G$ on the resolution $\Delta p$ of the applied pressure and on the typical spatial resolution $\Delta x$ in determining the characteristic points that are used to characterize the particle shape. The spatial resolution also leads to an error in the determination of the taper angle $\alpha$, which we estimate as $\Delta \alpha \approx \frac{\Delta x}{L_\mathrm{band}}$. The propagation of these errors leads to an error in the determination of the compressive modulus as $\Delta K = \frac{dK}{dp} \Delta p + \frac{dK}{d\epsilon_r} \frac{d \epsilon_r}{dx} \Delta x + \frac{dK}{d\epsilon_z} \frac{d \epsilon_z}{dx} \Delta x + \frac{dK}{d\alpha} \Delta \alpha$. In our experiments we estimate that the typical positional error is $\Delta x \approx$ 1 pixel, where typically $L_\mathrm{band} \approx 2 R_\mathrm{band} \approx \mathrm{\ 100\ pixels}$ and the accuracy of the pressure is $\frac{\Delta p}{p} \approx 5\%$. The estimated relative error in $K$ is then $\frac{\Delta K}{K} \approx \frac{\Delta p}{p} + \left( \frac{5}{2\epsilon_r} +1\right) \frac{\Delta x}{L_\mathrm{band} }$; for a typical compression $2\epsilon_r + \epsilon_z \approx 10\%$, $\Delta x / L_\mathrm{band} = 1\%$ and $\Delta p / p = 5\%$ we find $\frac{\Delta K}{K} \approx 30\%$.
Additional, systematic errors may come from friction between the particle and the glass walls, which has been neglected in our data analysis. 

For the hydrogel systems studied here, we observe a linear dependence of the stress on the strain, even at relatively large deformations. However, for many soft objects nonlinear behavior such as strain stiffening would be expected; given a sufficient range of applied stresses we would expect to be able to directly characterize this behavior as well.

To be able to further check the validity of our method, we perform macroscopic mechanical tests on macroscopic hydrogel samples, where in the synthesis we use the same monomer, cross-linker and initiator as used in our soft microgel particles. We cast the samples in petri dishes to obtain disc-shaped gels with a thickness of $\approx\ 5\ \mathrm{mm}$. We make such discs with a range of polymer concentrations, keeping the ratio of cross-linker to monomer constant for all samples.

We perform compressive tests on these samples using the normal force transducer of a strain-controlled rheometer (Rheometrics, ARES LS). The samples are allowed to swell in water to their equilibrium size and the compression tests are performed in water between two parallel steel plates by slowly decreasing the distance between the plates at a speed of typically $0.1 \ \mathrm{mm/min}$, while simultaneously measuring the normal force exerted on the plates by the compressing microgel.

For a poroelastic material such as the hydrogels studied here, sample size must have a significant influence on the relaxation time of the material; we therefore aim at characterizing only the elastic behavior at time scales longer than the relaxation time of the material. We ensure that we are measuring in this slow deformation regime by first compressing the sample and then extending it again. At slow enough rates of strain the amount of syneresis between the two curves should be minimal; indeed, at a rate of deformation of $0.1 \ \mathrm{mm/min}$ the curves for increasing and decreasing strain deformation almost overlap, as shown for a typical experiment in Fig.\ref{Mechanics_vs_cp}(B);  this indicates that the rate of deformation is slow enough to adequately suppress viscous contributions to the response.

In these bulk measurements we are not characterizing the full elastic behavior, as we only measure the response to a uniaxial compression, while leaving the sample free to expand or contract along the other axes; this yields the Young's modulus $E$ of the material. Thus, in this case as well we assume the absence of static friction between the sample and the walls of the compression plates. To compare the results from our capillary micromechanics measurements to the macroscopic tests, we also express them in terms of the Young's modulus, $E$, which is related to $K$ and $G$ via\cite{Hunter_mechanics_book}
\begin{equation}
E = \frac{9KG}{3K+G} ,
\end{equation}
and plot them in Fig.\ref{Mechanics_vs_cp}(C) as a function of the internal polymer concentration of the particles in their swollen state, $c_p V_o/V$. This accounts for the change of polymer concentration upon swelling of the particles, with $c_\mathrm{p}$ the polymer concentration during the initial drop formation stage, $V_0$ the volume of the drops, and $V$ the volume of the cross-linked particles after swelling in water.
Within our experimental error the magnitude of the Young's modulus is in good agreement with the macroscopic measurements at all polymer concentrations studied. Moreover, the dependence of $E$ on polymer concentration is well described by a power law behavior $E \propto (c_\mathrm{p} V_0/V)^\nu$ for both the macroscopic and the microscopic measurement, where the exponent $\nu \approx 2.9$ is identical in both cases. A power-law behavior is frequently observed in measurements on microgels and hydrogel systems\cite{geissler1988compressional,horkay1982studies,cohen2003characterization} and is also predicted theoretically from scaling arguments\cite{horkay1982studies,deGennes1979scaling}. However, the magnitude of the moduli found in the capillary measurements are consistently higher than those measured in the bulk measurements. This difference could be due to a not completely isotropic morphology of the particles - during their synthesis in the microfluidic device the initiator is flown in the outer phase, which could lead to a higher polymer concentration and/or cross-link density near the surface of the particles, which in turn could increase their overall elastic properties. Another reason for the higher values found in the capillary measurements could be the occurence of static friction in the capillary measurements, which we have neglected in our analysis.
Nevertheless our results indicate that the capillary micromechanics method provides reasonable results for the mechanics of soft particles at microscopic length scales.

\subsection*{Conclusions}
We have developed and validated a new technique for the characterization of soft objects, which we term \emph{capillary micromechanics}. Our approach enables the quantification of the full elastic response of soft particles, expressed in terms of the compressive modulus $K$ and the shear modulus $G$ from a single experiment. This approach enables us to derive the average properties of an entire soft object, while other methods such as atomic force microscopy (AFM) are more suited to characterize the localized response at nanometer length scales. For inhomogeneous soft objects such as biological cells it is not straightforward to extract information on the elastic properties of the whole object from such localized measurements. Our method should thus be important for the characterization of such systems, in combination with more localized methods such as AFM\cite{AFM_microgels_Lyon_2007,Matzelle:2003p6075,AFM_Rotsch_2000,AFM_Hassan_1998} or micropipette aspiration\cite{micropipette_aspiration_Kwok_Evans}.
Due to its simplicity and broad range of accessible elastic behaviors we expect this approach to be applicable to a wide range of soft and biological materials.

\subsection*{Acknowledgements}
We thank Jeremy Agresti for help with the microfluidic setup. This work was supported in part by the NSF (DMR-0602684), the Harvard MRSEC (DMR-0820484), and by grants from Italy (FIRB project RBIN045NMB and the Apulia Regional Strategic Project PS144).

\bibliographystyle{rsc}

\bibliography{squeezing}

\providecommand*{\mcitethebibliography}{\thebibliography}
\csname @ifundefined\endcsname{endmcitethebibliography}
{\let\endmcitethebibliography\endthebibliography}{}
\begin{mcitethebibliography}{30}
\providecommand*{\natexlab}[1]{#1}
\providecommand*{\mciteSetBstSublistMode}[1]{}
\providecommand*{\mciteSetBstMaxWidthForm}[2]{}
\providecommand*{\mciteBstWouldAddEndPuncttrue}
  {\def\EndOfBibitem{\unskip.}}
\providecommand*{\mciteBstWouldAddEndPunctfalse}
  {\let\EndOfBibitem\relax}
\providecommand*{\mciteSetBstMidEndSepPunct}[3]{}
\providecommand*{\mciteSetBstSublistLabelBeginEnd}[3]{}
\providecommand*{\EndOfBibitem}{}
\mciteSetBstSublistMode{f}
\mciteSetBstMaxWidthForm{subitem}
{(\emph{\alph{mcitesubitemcount}})}
\mciteSetBstSublistLabelBeginEnd{\mcitemaxwidthsubitemform\space}
{\relax}{\relax}

\bibitem[Malmsten(2006)]{malmsten2006soft}
M.~Malmsten, \emph{Soft Matter}, 2006, \textbf{2}, 760--769\relax
\mciteBstWouldAddEndPuncttrue
\mciteSetBstMidEndSepPunct{\mcitedefaultmidpunct}
{\mcitedefaultendpunct}{\mcitedefaultseppunct}\relax
\EndOfBibitem
\bibitem[Stokes and Frith(2008)]{stokes2008rheology}
J.~Stokes and W.~Frith, \emph{Soft Matter}, 2008, \textbf{4}, 1133--1140\relax
\mciteBstWouldAddEndPuncttrue
\mciteSetBstMidEndSepPunct{\mcitedefaultmidpunct}
{\mcitedefaultendpunct}{\mcitedefaultseppunct}\relax
\EndOfBibitem
\bibitem[Jeong and Gutowska(2002)]{jeong2002lessons}
B.~Jeong and A.~Gutowska, \emph{Trends in Biotechnology}, 2002, \textbf{20},
  305--311\relax
\mciteBstWouldAddEndPuncttrue
\mciteSetBstMidEndSepPunct{\mcitedefaultmidpunct}
{\mcitedefaultendpunct}{\mcitedefaultseppunct}\relax
\EndOfBibitem
\bibitem[Kasza \emph{et~al.}(2007)Kasza, Rowat, Liu, Angelini, Brangwynne,
  Koenderink, and Weitz]{kasza2007cell}
K.~Kasza, A.~Rowat, J.~Liu, T.~Angelini, C.~Brangwynne, G.~Koenderink and
  D.~Weitz, \emph{Current opinion in cell biology}, 2007, \textbf{19},
  101--107\relax
\mciteBstWouldAddEndPuncttrue
\mciteSetBstMidEndSepPunct{\mcitedefaultmidpunct}
{\mcitedefaultendpunct}{\mcitedefaultseppunct}\relax
\EndOfBibitem
\bibitem[Adams \emph{et~al.}(2004)Adams, Frith, and Stokes]{adams2004influence}
S.~Adams, W.~Frith and J.~Stokes, \emph{Journal of Rheology}, 2004,
  \textbf{48}, 1195\relax
\mciteBstWouldAddEndPuncttrue
\mciteSetBstMidEndSepPunct{\mcitedefaultmidpunct}
{\mcitedefaultendpunct}{\mcitedefaultseppunct}\relax
\EndOfBibitem
\bibitem[Borrega \emph{et~al.}(1999)Borrega, Cloitre, Betremieux, Ernst, and
  Leibler]{Borrega:1999p6058}
R.~Borrega, M.~Cloitre, I.~Betremieux, B.~Ernst and L.~Leibler, \emph{EPL
  (Europhysics Letters)}, 1999, \textbf{47}, 729--735\relax
\mciteBstWouldAddEndPuncttrue
\mciteSetBstMidEndSepPunct{\mcitedefaultmidpunct}
{\mcitedefaultendpunct}{\mcitedefaultseppunct}\relax
\EndOfBibitem
\bibitem[Ketz \emph{et~al.}(1988)Ketz, Prud'homme, and
  Graessley]{Ketz:1988p2084}
R.~Ketz, R.~Prud'homme and W.~Graessley, \emph{Rheologica Acta}, 1988,
  \textbf{27}, 531--539\relax
\mciteBstWouldAddEndPuncttrue
\mciteSetBstMidEndSepPunct{\mcitedefaultmidpunct}
{\mcitedefaultendpunct}{\mcitedefaultseppunct}\relax
\EndOfBibitem
\bibitem[Mattsson \emph{et~al.}(2009)Mattsson, Wyss, Fernandez-Nieves,
  Miyazaki, Hu, Reichman, and Weitz]{Mattsson:2009p6046}
J.~Mattsson, H.~M. Wyss, A.~Fernandez-Nieves, K.~Miyazaki, Z.~Hu, D.~R.
  Reichman and D.~A. Weitz, \emph{Nature}, 2009, \textbf{462}, 83--86\relax
\mciteBstWouldAddEndPuncttrue
\mciteSetBstMidEndSepPunct{\mcitedefaultmidpunct}
{\mcitedefaultendpunct}{\mcitedefaultseppunct}\relax
\EndOfBibitem
\bibitem[Saunders and Vincent(1999)]{Saunders:1999p2086}
B.~Saunders and B.~Vincent, \emph{Advances in Colloid and Interface Science},
  1999, \textbf{80}, 1--25\relax
\mciteBstWouldAddEndPuncttrue
\mciteSetBstMidEndSepPunct{\mcitedefaultmidpunct}
{\mcitedefaultendpunct}{\mcitedefaultseppunct}\relax
\EndOfBibitem
\bibitem[Wiedemair \emph{et~al.}(2007)Wiedemair, Serpe, Kim, Masson, Lyon,
  Mizaikoff, and Kranz]{AFM_microgels_Lyon_2007}
J.~Wiedemair, M.~J. Serpe, J.~Kim, J.-F. Masson, L.~A. Lyon, B.~Mizaikoff and
  C.~Kranz, \emph{Langmuir}, 2007, \textbf{23}, 130--137\relax
\mciteBstWouldAddEndPuncttrue
\mciteSetBstMidEndSepPunct{\mcitedefaultmidpunct}
{\mcitedefaultendpunct}{\mcitedefaultseppunct}\relax
\EndOfBibitem
\bibitem[Matzelle \emph{et~al.}(2003)Matzelle, Geuskens, and
  Kruse]{Matzelle:2003p6075}
T.~Matzelle, G.~Geuskens and N.~Kruse, \emph{Macromolecules}, 2003,
  \textbf{36}, 2926--2931\relax
\mciteBstWouldAddEndPuncttrue
\mciteSetBstMidEndSepPunct{\mcitedefaultmidpunct}
{\mcitedefaultendpunct}{\mcitedefaultseppunct}\relax
\EndOfBibitem
\bibitem[Rotsch and Radmacher({2000})]{AFM_Rotsch_2000}
C.~Rotsch and M.~Radmacher, \emph{Biophysical Journal}, {2000}, \textbf{{78}},
  {520--535}\relax
\mciteBstWouldAddEndPuncttrue
\mciteSetBstMidEndSepPunct{\mcitedefaultmidpunct}
{\mcitedefaultendpunct}{\mcitedefaultseppunct}\relax
\EndOfBibitem
\bibitem[A-Hassan \emph{et~al.}(1998)A-Hassan, Heinz, Antonik, D'Costa,
  Nageswaran, Schoenenberger, and Hoh]{AFM_Hassan_1998}
E.~A-Hassan, W.~Heinz, M.~Antonik, N.~D'Costa, S.~Nageswaran, C.~Schoenenberger
  and J.~Hoh, \emph{Biophysical Journal}, 1998, \textbf{74}, 1564--1578\relax
\mciteBstWouldAddEndPuncttrue
\mciteSetBstMidEndSepPunct{\mcitedefaultmidpunct}
{\mcitedefaultendpunct}{\mcitedefaultseppunct}\relax
\EndOfBibitem
\bibitem[Tagit \emph{et~al.}(2008)Tagit, Tomczak, and Vancso]{tagit2008probing}
O.~Tagit, N.~Tomczak and G.~Vancso, \emph{Small}, 2008, \textbf{4}, 119\relax
\mciteBstWouldAddEndPuncttrue
\mciteSetBstMidEndSepPunct{\mcitedefaultmidpunct}
{\mcitedefaultendpunct}{\mcitedefaultseppunct}\relax
\EndOfBibitem
\bibitem[Kwok and Evans(1981)]{micropipette_aspiration_Kwok_Evans}
R.~Kwok and E.~Evans, \emph{Biophysical Journal}, 1981, \textbf{35},
  637--652\relax
\mciteBstWouldAddEndPuncttrue
\mciteSetBstMidEndSepPunct{\mcitedefaultmidpunct}
{\mcitedefaultendpunct}{\mcitedefaultseppunct}\relax
\EndOfBibitem
\bibitem[Xia and Whitesides(1998)]{Xia_Whiteside_Angewandte_1998}
Y.~Xia and G.~M. Whitesides, \emph{Angewandte Chemie International Edition},
  1998, \textbf{37}, 550--575\relax
\mciteBstWouldAddEndPuncttrue
\mciteSetBstMidEndSepPunct{\mcitedefaultmidpunct}
{\mcitedefaultendpunct}{\mcitedefaultseppunct}\relax
\EndOfBibitem
\bibitem[Shah \emph{et~al.}(2008)Shah, Kim, Agresti, Weitz, and
  Chu]{shah2008fabrication}
R.~Shah, J.~Kim, J.~Agresti, D.~Weitz and L.~Chu, \emph{Soft Matter}, 2008,
  \textbf{4}, 2303--2309\relax
\mciteBstWouldAddEndPuncttrue
\mciteSetBstMidEndSepPunct{\mcitedefaultmidpunct}
{\mcitedefaultendpunct}{\mcitedefaultseppunct}\relax
\EndOfBibitem
\bibitem[Pelton(2000)]{pelton2000temperature}
R.~Pelton, \emph{Advances in colloid and interface science}, 2000, \textbf{85},
  1--33\relax
\mciteBstWouldAddEndPuncttrue
\mciteSetBstMidEndSepPunct{\mcitedefaultmidpunct}
{\mcitedefaultendpunct}{\mcitedefaultseppunct}\relax
\EndOfBibitem
\bibitem[Squires and Quake(2005)]{squires2005microfluidics}
T.~Squires and S.~Quake, \emph{Reviews of modern physics}, 2005, \textbf{77},
  977--1026\relax
\mciteBstWouldAddEndPuncttrue
\mciteSetBstMidEndSepPunct{\mcitedefaultmidpunct}
{\mcitedefaultendpunct}{\mcitedefaultseppunct}\relax
\EndOfBibitem
\bibitem[Delamarche \emph{et~al.}(1998)Delamarche, Bernard, Schmid, Bietsch,
  Michel, and Biebuyck]{delamarche1998microfluidic}
E.~Delamarche, A.~Bernard, H.~Schmid, A.~Bietsch, B.~Michel and H.~Biebuyck,
  \emph{J. Am. Chem. Soc}, 1998, \textbf{120}, 500--508\relax
\mciteBstWouldAddEndPuncttrue
\mciteSetBstMidEndSepPunct{\mcitedefaultmidpunct}
{\mcitedefaultendpunct}{\mcitedefaultseppunct}\relax
\EndOfBibitem
\bibitem[Hou \emph{et~al.}(2009)Hou, Li, Lee, Kumar, Ong, and
  Lim]{hou2009deformability}
H.~Hou, Q.~Li, G.~Lee, A.~Kumar, C.~Ong and C.~Lim, \emph{Biomedical
  Microdevices}, 2009, \textbf{11}, 557--564\relax
\mciteBstWouldAddEndPuncttrue
\mciteSetBstMidEndSepPunct{\mcitedefaultmidpunct}
{\mcitedefaultendpunct}{\mcitedefaultseppunct}\relax
\EndOfBibitem
\bibitem[Arima and Iwata(2007)]{arima2007effects}
Y.~Arima and H.~Iwata, \emph{Journal of Materials Chemistry}, 2007,
  \textbf{17}, 4079--4087\relax
\mciteBstWouldAddEndPuncttrue
\mciteSetBstMidEndSepPunct{\mcitedefaultmidpunct}
{\mcitedefaultendpunct}{\mcitedefaultseppunct}\relax
\EndOfBibitem
\bibitem[Hunter(1976)]{Hunter_mechanics_book}
S.~Hunter, \emph{Mechanics of Continuous Media}, Ellis Horwood, 2nd edn.,
  1976\relax
\mciteBstWouldAddEndPuncttrue
\mciteSetBstMidEndSepPunct{\mcitedefaultmidpunct}
{\mcitedefaultendpunct}{\mcitedefaultseppunct}\relax
\EndOfBibitem
\bibitem[Macosko(1994)]{Macosko_Rheology_book}
C.~Macosko, \emph{Rheology: principles, measurements, and applications},
  Wiley-VCH, 1994\relax
\mciteBstWouldAddEndPuncttrue
\mciteSetBstMidEndSepPunct{\mcitedefaultmidpunct}
{\mcitedefaultendpunct}{\mcitedefaultseppunct}\relax
\EndOfBibitem
\bibitem[Li \emph{et~al.}(1993)Li, Hu, and Li]{Li:1993p6077}
Y.~Li, Z.~Hu and C.~Li, \emph{Journal of Applied Polymer Science}, 1993,
  \textbf{50}, 1107--1111\relax
\mciteBstWouldAddEndPuncttrue
\mciteSetBstMidEndSepPunct{\mcitedefaultmidpunct}
{\mcitedefaultendpunct}{\mcitedefaultseppunct}\relax
\EndOfBibitem
\bibitem[Geissler \emph{et~al.}(1988)Geissler, Hecht, Horkay, and
  Zrinyi]{geissler1988compressional}
E.~Geissler, A.~Hecht, F.~Horkay and M.~Zrinyi, \emph{Macromolecules}, 1988,
  \textbf{21}, 2594--2599\relax
\mciteBstWouldAddEndPuncttrue
\mciteSetBstMidEndSepPunct{\mcitedefaultmidpunct}
{\mcitedefaultendpunct}{\mcitedefaultseppunct}\relax
\EndOfBibitem
\bibitem[Horkay and Zrinyi(1982)]{horkay1982studies}
F.~Horkay and M.~Zrinyi, \emph{Macromolecules}, 1982, \textbf{15},
  1306--1310\relax
\mciteBstWouldAddEndPuncttrue
\mciteSetBstMidEndSepPunct{\mcitedefaultmidpunct}
{\mcitedefaultendpunct}{\mcitedefaultseppunct}\relax
\EndOfBibitem
\bibitem[Cohen \emph{et~al.}(2003)Cohen, Ramon, Kopelman, and
  Mizrahi]{cohen2003characterization}
Y.~Cohen, O.~Ramon, I.~Kopelman and S.~Mizrahi, \emph{Journal of Polymer
  Science Part B: Polymer Physics}, 2003, \textbf{30}, 1055--1067\relax
\mciteBstWouldAddEndPuncttrue
\mciteSetBstMidEndSepPunct{\mcitedefaultmidpunct}
{\mcitedefaultendpunct}{\mcitedefaultseppunct}\relax
\EndOfBibitem
\bibitem[de~Gennes(1979)]{deGennes1979scaling}
P.~de~Gennes, \emph{Scaling concepts in polymer physics}, Cornell Univ Pr,
  1979\relax
\mciteBstWouldAddEndPuncttrue
\mciteSetBstMidEndSepPunct{\mcitedefaultmidpunct}
{\mcitedefaultendpunct}{\mcitedefaultseppunct}\relax
\EndOfBibitem
\end{mcitethebibliography}

\end{document}